\documentclass[aps,prl,twocolumn,showpacs, superscriptaddress]{revtex4-1}

\usepackage{graphicx}

\begin{document}


\title{Quantifying, characterizing and controlling information flow in ultracold atomic gases}



\author{P. Haikka}
\email[]{pmehai@utu.fi}
\homepage[]{www.openq.fi}
\affiliation{Turku Center for Quantum Physics, Department of Physics and Astronomy, University of Turku, FIN-20014 Turku, Finland}
\author{S. McEndoo}
\affiliation{Turku Center for Quantum Physics, Department of Physics and Astronomy, University of Turku, FIN-20014 Turku, Finland}
\affiliation{SUPA, EPS/Physics, Heriot-Watt University, Edinburgh, EH144AS, UK}
\author{G. De Chiara}
\affiliation{F\'isica Te\`orica: Informaci\'o i Fen\`{o}mens  Qu\`antics, Universitat Aut\`{o}noma de Barcelona, E-08193 Bellaterra, Spain}
\author{G. M. Palma}
\affiliation{NEST Istituto Nanoscienze-CNR and Dipartimento di Fisica, Universit\`a degli Studi di Palermo, via Archirafi 36, I-90123 Palermo, Italy}
\author{S. Maniscalco} 
\affiliation{Turku Center for Quantum Physics, Department of Physics and Astronomy, University of Turku, FIN-20014 Turku, Finland}
\affiliation{SUPA, EPS/Physics, Heriot-Watt University, Edinburgh, EH144AS, UK}
\email[]{smanis@utu.fi} \homepage[]{www.openq.fi}


\date{\today}

\begin{abstract}
We study quantum information flow in a model comprising of an impurity qubit immersed in a Bose-Einstein condensed reservoir. We demonstrate how information flux between the qubit and the condensate can be manipulated by engineering the ultracold reservoir within experimentally realistic limits. We place a particular emphasis on non-Markovian dynamics, characterized by a reversed flow of information from the background gas to the qubit and identify a controllable crossover between Markovian and non-Markovian dynamics in the parameter space of the model.
\end{abstract}


\maketitle

{\sl Introduction-} In the last decades high precision control of ultracold atomic gases has allowed the realization of beautiful experiments unveiling fundamental phenomena in the physics of many-body quantum systems at low temperatures. Key examples are the observation of Anderson localization \cite{anderson}, the superfluid-Mott insulator transition \cite{superfluid-mott}, the creation of Tonks-Girardeau gases \cite{TG}, and the atom laser \cite{atomlaser}, just to mention a few.\\
More recently, hybrid systems composed of quantum dots, single trapped ions and optical lattices coupled to Bose-Einstein condensates (BECs) have been studied both theoretically and experimentally \cite{hybrid systems}. These systems are studied in the framework of open quantum systems \cite{breuer&petruccione}, effectively described as one or more two-level systems (qubits) interacting with a reservoir consisting of the ultracold gas. The possibility of manipulating crucial parameters of the reservoir, such as the scattering length \cite{feshbach}, combined with the continuous improvements in quantum control of qubits highlights the enormous potential of hybrid systems as quantum simulators of both condensed matter models and open quantum systems. \\
Lately, the open systems community has given substantial interest to the dynamics of quantum information flow due to several proposals to link it to the division of quantum processes into Markovian and non-Markovian ones \cite{BLP, RHP, Kossakowski, Fisher}. The latter ones have been defined as processes where an open system recovers some previously lost information and therefore temporarily combats the destructive effect of the environment as a sink for quantum properties. Such processes are very desirable from a quantum information processing point of view, since they increase the operational time of qubits \cite{nilsen}. Sophisticated control over detrimental environmental effects enabled by, for example, hybrid quantum systems will be very beneficial for state-of-the-art quantum processors.\\
Moreover, understanding of the decoherence processes in hybrid systems gives insight into the basic physical mechanisms ruling the interaction of the qubit and the cold atomic cloud surrounding it. It is also a key pre-requisite to the realization of quantum simulators since it allows for the identification of the relevant parameters for on-demand control of the exchange of quantum information between a qubit and its ultracold environment.\\
We present here the first characterization of non-Markovian effects in the context of ultracold gases. More specifically, inspired by a model introduced in Ref. \cite{original}, we consider an impurity atom in a double well potential, i.e., an effective qubit system, interacting with a BEC environment. We study information flow using the non-Markovianity quantifier presented in Ref. \cite{BLP} and show how information flux can be manipulated by experimentally achievable means such as changing the scattering length, the effective dimension of the background gas or the trapping geometry of the qubit. We find a crossover between Markovian and non-Markovian dynamics in the parameter space of the model and uncover the physical mechanisms at the root of non-Markovian phenomena induced by the ultracold background gas. Our findings pave the way to the realization of quantum simulators for non-Markovian open quantum system models with ultracold atomic gases.

{\sl The model-} We consider an impurity atom trapped in a deep double well potential $V_A(\mathbf{ r})$. The impurity atom forms a qubit system with the two qubit states represented by the occupation of the impurity atom in the left or the right well, $|L\rangle$ and $|R\rangle$ respectively. The impurity atom couples to a bosonic background gas B trapped in a shallow potential $V_B(\mathbf{ r})$, which forms a Bose-Einstein condensed environment for the qubit system. The Hamiltonian for this system, derived in Ref. \cite{original}, is
\begin{equation}
\label{Hamiltonian}
H=\sum_\mathbf{ k} E_\mathbf{ k} c_\mathbf{ k}^\dagger c_\mathbf{ k}+\sigma_z\sum_\mathbf{ k}(g_\mathbf{ k}c_\mathbf{ k}^ \dagger +g_\mathbf{ k}^*c_\mathbf{ k})+\sum_\mathbf{ k}(\xi_\mathbf{ k}c_\mathbf{ k}^ \dagger +\xi_\mathbf{ k}^*c_\mathbf{ k}),
\end{equation}
where $\sigma_z=|R\rangle\langle R|-|L\rangle\langle L|$ and $E_\mathbf{ k}=\sqrt{\epsilon_\mathbf{ k}[\epsilon_\mathbf{ k}+2 g_B^{(D)}n_D]}$ is the energy of $k$-th Bogoliubov mode $c_\mathbf{ k}$ of the condensate with boson-boson coupling frequency $g_B^{(D)}$ and condensate density  $n_D$. $D$ denotes the effective dimension of the environment. The energy of a free mode is $\epsilon_\mathbf{ k}=\hbar^2k^2/(2m_B)$ where $k=|\mathbf{ k}|$ and $m_B$ is the mass of a background gas particle. Furthermore, $g_\mathbf{k}$ and $\xi_\mathbf{k}$ are coupling constants that depend on the spatial form of the states $|L\rangle$ and $|R\rangle$ and on the shape of the Bogoliubov modes. Their specific form is elaborated in Ref. \cite{original}.\\
When the background gas is at zero temperature the reduced dynamics of the impurity atom is captured by the following time-local master equation (ME):
\begin{equation}
 \label{TCL2}
\frac{d\rho(t)}{dt}=\Lambda(t)[\sigma_z,\rho]+\gamma(t)[\sigma_z\rho(t)\sigma_z-\frac{1}{2}\{\sigma_z\sigma_z,\rho(t)\}].
\end{equation}
Quantity $\Lambda(t)$ renormalizes the energy of the qubit but has no qualitative effect on the dissipative dynamics. Instead in this work we are interested in the decay rate
\begin{eqnarray}
\label{decoherence}
\gamma(t)=&&\frac{4 g_{AB}^2n_0}{\hbar}\int \frac{d\mathbf{ k}\sin^2(\mathbf{k}\cdot\mathbf{L})}{(2\pi)^D} \frac{\sin(E_\mathbf{ k} t/\hbar)}{\epsilon_\mathbf{ k}+2g_B^{(D)}n_D}e^{-k^2\tau^2/2},\nonumber\\
\end{eqnarray}
where $g_{AB}$ is the impurity-boson coupling frequency, $\tau$ is a trap parameter, and $\mathbf{ L}$ is half the distance between the two wells of the double well potential.\\
We have derived ME (\ref{TCL2}) using the time-convolutionless projection operator technique to second order in the coupling constant $g_{AB}$ \cite{breuer&petruccione}. Remarkably, in this case the second order ME describes the reduced dynamics exactly \cite{exact}. Solving the ME reveals that the impurity atom dephases without exchanging energy with the background gas. More precisely, $\rho_{ii}(t)=\rho_{ii}(0)$ and $\rho_{ij}(t)=e^{-\Gamma(t)}\rho_{ij}(0)$ when $i\neq j$, where $\rho_{ij}=\langle i|\rho|j\rangle$, $i,j=R,L$. The decoherence function $\Gamma(t)=\int_0^tds\,\gamma(s)$ coincides with that derived in Ref. \cite{original}, however here we wish to stress the connection between the decay rate and the non-Markovian features. The authors of Ref. \cite{original} discovered situations when the decoherence function $\Gamma(t)$ is non-monotonic and conjectured that this is due to non-Markovian effects in the reduced dynamics. Already the form of the ME (\ref{TCL2}) supports this intuition; the theory of non-Markovian quantum jumps has shown that there is a profound connection between non-Markovian effects and negative regions of the decay rates of Lindblad-structured MEs as the one of Eq. (\ref{TCL2}) \cite{nmqj}. In the following we confirm that this is indeed true for this model and, moreover, expose the physical mechanisms at the root of this non-Markovian phenomena.

{\sl Non-Markovianity measure-}  Breuer, Laine and Piilo (BLP) have proposed a rigorous definition for non-Markovianity of a quantum channel $\Phi$ based on the dynamics of the so-called information flux $\sigma(t)=dD[\rho_1(t), \rho_2(t)]/dt$ \cite{BLP}. This is the temporal change in the distinguishability $D[\rho_1(t),\rho_2(t)]=\frac{1}{2}||\rho_1(t)-\rho_2(t)||_1$ of two evolving quantum states $\rho_{1,2}(t)=\Phi(t)\rho_{1,2}(0)$ as measured by the trace distance. Negative information flux describes information leaking from the system to its environment and it is associated to Markovian dynamics. Instead, if it is possible to find a pair of states $\rho_{1,2}(0)$ for which the information flux is positive for some interval of time, that is, the system regains some of the previously lost information, then process $\Phi$ is considered non-Markovian. The amount of non-Markovianity is defined to be the maximal amount of information that the system may recover from its environment, formally $\mathcal{N}_{BLP}=\max_{\rho_{1,2}}\int_{\sigma>0}ds\sigma(s)$.\\
For the model studied in this Letter we find that $\sigma(t)>0$ if and only if $\gamma(t)<0$, that is, the process is non-Markovian precisely when the decay rate can take temporarily negative values. Within experimentally relevant values of the physical parameters we have discovered at most a single time-interval $t\in[a,b]$ when the decay rate is negative and information flows back to the system after an initial period of information loss. Therefore, instead of using the original measure faithfully and quantifying non-Markovianity as the maximal amount of information that the system may recover, we introduce a normalized quantity that reveals the maximal {\sl  fraction} of the previously lost information that the system can recover:
\begin{equation}
\label{NM}
\mathcal{N}=\max_{\rho_{1,2}}\frac{D[\rho_1(b),\rho_2(b)]-D[\rho_1(a),\rho_2(a)]}{D[\rho_1(0),\rho_2(0)]-D[\rho_1(a),\rho_2(a)]}.
\end{equation}
Unlike $\mathcal{N}_{BLP}$, the modified quantifier $\mathcal{N}$ is bounded between zero (system only leaks information) and one (system regains all previously lost information) and is therefore more meaningful as a number. We have confirmed numerically that in the relevant case of dephasing noise the above quantity is maximized for the same pair of initial states that maximize $\mathcal{N}_{BLP}$. These are the states whose Bloch vectors lie on the opposite sides of the equator of the Bloch sphere \cite{analyticalN}. Using these states in the general expression of Eq. (\ref{NM}) we find the analytic expression of the non-Markovianity measure for a dephasing qubit to be
\begin{equation}
\label{ }
\mathcal{N}_{deph}=\frac{e^{-\Gamma(b)}-e^{-\Gamma(a)}}{e^{-\Gamma(0)}-e^{-\Gamma(a)}},\quad \Gamma(t)=\int_0^tds\,\gamma(s).
\end{equation}
We are now ready to study how changes in the background scattering length and in the dimensionality of the BEC affect the dynamics of information flow.

{\sl Three-dimensional BEC-} As a first step we consider a 3D background BEC with equal confinement of the background gas in all directions. We consider a $^{87}$Rb-condensate of density $n_3=n_0=10^{20}$m$^{-3}$ and $^{23}$Na impurity atoms trapped in an optical lattice with lattice wavelength $\lambda=600$nm and trap parameter $\tau=45$nm. The impurity-boson coupling is $g_{AB}=2\pi\hbar^2a_{AB}/m_{AB}$, where $m_{AB}=m_Am_B/(m_A+m_B)$ and $m_A$ and $m_B$ are the masses of the impurity atoms and the bosons, respectively, and $a_{AB}=55\,a_0$, where $a_0$ is the Bohr radius. Similarly the boson-boson coupling frequency is $g_B^{3D}=4\pi\hbar^2a_{B}/m_B$ but now we assume that the s-wave scattering length of the background gas can be tuned from its natural value $a_B=a_{Rb}\approx5.3$nm via Feshbach resonances. We explore a range of values of $a_B$ consistent with the assumption of dilute gas and with the regime of weakly interacting gases. The latter is a stronger condition, requiring $\sqrt{a_B^3 n_0}\ll1$. As a consequence, we can tune the scattering length up to a maximum value given by $a_B\approx3\,a_{Rb}$.\\ 
Figure. \ref{N(ab,L)} shows the non-Markovianity measure $\mathcal{N}_{deph}$ as a function of $a_B$ for three different values of the well separation $L$. Increasing $L$ magnifies the fraction of recovered information flow due to the increased ability of the condensate to resolve the qubit system. Similarly, non-Markovian effects are amplified for stronger interaction of the condensate. However we find that the scattering length alone plays a crucial role in the emergence of non-Markovian reservoir memory effects. When the background gas is free or very weakly interacting, $0\leq a_B\leq a_B^{crit}\approx 0.034\, a_{Rb}$, the qubit only leaks information to the BEC environment. Instead for a strong enough interaction strength of the background gas, $a_B> a_B^{crit}$, the condensate can take on the role of information storage and feed some information back to the qubit. This result holds for any value of $L$. This finding challenges the conclusion of Ref. \cite{original}, where the scattering length dependent Markovian/non-Markovian crossover was only attributed to the 1D case. We have discovered that the situation is indeed more subtle and we will show next that the crossover point exists in all three dimensions.

\begin{figure}[t]
\includegraphics[width=1\linewidth]{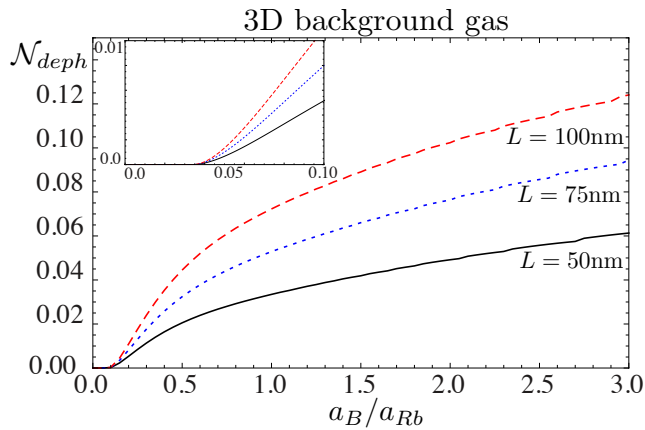}
\caption{(Color online) Non-Markovianity measure $\mathcal{N}_{deph}$ as a function of the scattering length of the background gas $a_B$ with values of well separation $L=50$nm (red dashed line), $L=75$nm (blue dotted line) and $L=100$nm (black solid line). The inset shows $\mathcal{N}_{deph}$ for very small values of the scattering lenght.}
\label{N(ab,L)}
\end{figure}

\begin{figure}[h]
\includegraphics[width=0.49\textwidth]{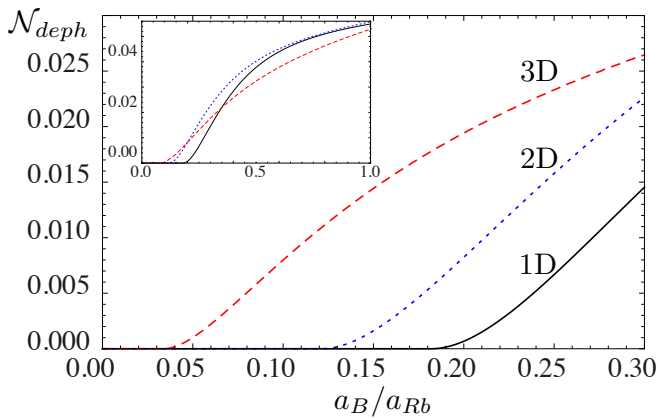}
\caption{(Color online) Non-Markovianity measure $\mathcal{N}_{deph}$ as a function of the scattering length of the background gas $a_B$ when the background gas is three dimensional (red dashed line), quasi-two dimensional (blue dotted line) and quasi-one dimensional (black solid line). The inset shows a longer range of the scattering length $a_B$.  In all figures the well separation is $L=75$nm.}
\label{fig2}
\end{figure}

{\sl Lower dimensions-} By a suitable modification of the potential of the condensate $V_B(\mathbf{r})$ we can create a quasi-2D background gas where the gas is trapped in a slightly anisotropic, pancake-shaped harmonic trap. Assuming that the scattering length is still much smaller than the axial length of the condensate, $a_B\ll a_z$, the coupling term is modified to $g^{2D}_B=\sqrt{8 \pi}\hbar^2 a_B/(m_B a_z)$ and the 2D condensate density is $n_2 = \sqrt{\pi} n_0 a_z$ \cite{BEC}. Within the limits of a dilute gas we can increase the scattering length up to $a_B\approx2\,a_{Rb}.$ The potential $V_B(\mathbf{r})$ can be also modified to create a cigar-shaped quasi-1D background gas with transversal width $a_{\bot}$. The consequent coupling is $g^{1D}_B =2\hbar^2a_B/(m_Ba_{\bot}^2)$ and the 1D density is $n_1=n_0 \pi a_{\bot}^2$, again provided that gas is weakly enough confined, $a_B\ll a_{\bot}$ \cite{bloch review}. In the quasi-1D regime diluteness of the gas allows at most $a_B\lesssim a_{Rb}$.\\
In Fig. \ref{fig2} we plot the non-Markovianity measure $\mathcal{N}_{deph}$ in the quasi-1D, quasi-2D and 3D cases. In lower dimensions we find the critical values $a_B^{crit,\text{2D}}\approx0.122a_{Rb}$ and $a_B^{crit,\text{1D}}\approx0.183a_{Rb}$. Clearly when the dimensionality of the background gas is lowered the cross-over value of the scattering length $a_B^{crit}$ increases. Crucially, as we remarked before, $a_B^{crit,\text{3D}}>0$ and therefore it is possible to create both Markovian and non-Markovian dynamical processes in all three dimensions.\\
We note here that the quantities we have chosen to vary, namely, the scattering length $a_B$ and the well separation $L$, are indeed the most relevant quantities for manipulating the information flow-back. The trap parameter $\tau$ determines the trapping frequency of the double well trap $V_A(\mathbf{r})$ and acts as a natural cut-off parameter in the decay rate of Eq. (\ref{decoherence}). As long as the double well trap is deep enough to prevent hopping between the two sites the particular value of $\tau$ has only a minor effect on information flow. It is also  clear from the form of the decay rate that the boson-impurity coupling $g_{AB}$ cannot affect the Markovian/non-Markovian crossover. Moreover, we have found that its value has negligible effect on the non-Markovianity quantifier $\mathcal{N}_{deph}$. Intuitively, it affects the amount of outgoing and incoming information almost equally but leaves their ratio unchanged. In order to explain the  key non-Markovian features in the dynamics of the qubit system when the dimensionality and the scattering length of the background gas vary, we need to take a closer look at the spectrum of the BEC reservoir.

{\sl Environmental spectrum-} The crossover between Markovian and non-Markovian processes is best understood in terms of the environmental spectrum $J(\omega)$. Consider the general dephasing model introduced by Palma, Suominen and Ekert describing qubit dynamics: $\rho_{ii}(t)=\rho_{ii}(0)$ and $\rho_{ij}(t)=e^{-\tilde{\Gamma}(t)}\rho_{ij}(0)$, where $\tilde{\Gamma}(t)\sim\int d\omega J(\omega)(1-\cos\omega t)/\omega$ \cite{dephasing}. The dynamical process is non-Markovian if and only if $\tilde{\gamma}(t)=d\tilde{\Gamma}(t)/dt<0$ for some interval of time. We assume an Ohmic-type spectrum $J(\omega)\sim\omega^s$ and recall the convention that the spectrum is sub-Ohmic when $s<1$, Ohmic when $s=1$ or super-Ohmic when $s>1$. Introducing an {\sl ad hoc} exponential cut-off so that $J(\omega)=\omega^s \exp\{-\omega^2/\omega_C^2\}$, where $\omega_C$ is the cut-off frequency, it is straightforward to show that the dynamics is non-Markovian when $s>s^{crit}=2$. Therefore, in a general setting, a qubit dephasing under the effect of either a sub-Ohmic or an Ohmic environment can only leak information to its environment. If the environment has a super-Ohmic spectrum the issue is less straightforward: only if the spectrum is sufficiently super-Ohmic with $s>s^{crit}$, information can flow back to the system from the environment.\\
The qubit in an ultracold bosonic environment considered in this work is a special case of the model above with $J(\omega)=\sum_{\mathbf{k}}|g_\mathbf{ k}|^2\delta(\hbar \omega-E_\mathbf{k})$. The reservoir spectrum $J(\omega)$ is very complex due to the complicated form of the coupling constant $g_\mathbf{k}$. However, it can be shown that in the case of a free background gas in one, two or three dimensions the spectrum is sub-Ohmic, Ohmic or super-Ohmic, respectively \cite{original}. The spectrum changes critically when one considers the boson-boson coupling quantified by the scattering length $a_B$. In this case increasing the scattering length effectively increases the value of $s$. Hence when we increase $a_B$ in the 1D case, the spectrum changes from sub-Ohmic to Ohmic to super-Ohmic and once a critical threshold of super-Ohmicity is reached the environment can feed information back to the system. In the 2D non-interacting case the spectrum is Ohmic and a weaker interaction is required to reach the crossover point $s^{crit}$, leading to $a_B^{crit,\text{2D}}<a_B^{crit,\text{1D}}$. Finally, in the 3D case the spectrum is already super-Ohmic in the non-interacting case, although not super-Ohmic enough to give rise to non-Markovian dynamics. Already a small increase in the scattering length modifies the spectrum so that the direction of information flow can be temporarily reversed.

{\sl Conclusion-} We have studied quantum information flux in an ultracold hybrid system of an impurity atom immersed in a BEC environment. We have shown explicitly how precise control of the ultracold background gas affects the spectrum felt by the qubit and therefore enables the manipulation of the qubit dynamics and the information flux. In particular, we have discovered experimentally accessible means to reach non-Markovian dynamical regimes, where the background gas may feed information back to the qubit instead of acting only as a sink for information. Such quantum reservoir engineering is fundamental for understanding decoherence processes in quantum information processing and, more specifically, for the realization of quantum simulators.

\begin{acknowledgments}
This work was supported by the Emil Aaltonen foundation, the Finnish Cultural foundation and the Spanish MICINN (Juan de la Cierva, FIS2008-01236 and QOIT-Consolider Ingenio 2010), Generalitat de Catalunya Grant No. 2005SGR-00343. We acknowledge Markus Cirone, Francesco Plastina and John Goold for useful discussions.
\end{acknowledgments}

\end{document}